\documentclass[aps,prl,twocolumn,superscriptaddress,floatfix]{revtex4-1}

\usepackage{graphicx}
\usepackage{amssymb}
\usepackage{epsfig}
\usepackage[dvipsnames]{xcolor}
\begin{document}

%\title{Interplay of timescales in the flow of glassy mixtures with large size ratio}
\title{Rheological response of a glass-forming liquid having large bidispersity}
\author{Vinay Vaibhav}
\affiliation{The Institute of Mathematical Sciences, CIT Campus, Taramani, Chennai 600113, India}
\affiliation{Homi Bhabha National Institute, Anushaktinagar, Mumbai 400094, India}
\author{J\"urgen Horbach}
\affiliation{Institut f\"ur Theoretische Physik II, Heinrich-Heine-Universit\"at
D\"usseldorf, Universit\"atsstra\ss e 1, 40225 D\"usseldorf, Germany}
\author{Pinaki Chaudhuri}
\affiliation{The Institute of Mathematical Sciences, CIT Campus, Taramani, Chennai 600113, India}
\affiliation{Homi Bhabha National Institute, Anushaktinagar, Mumbai 400094, India}

%%%%%%%%%%%%%%%%%%%%%%%%%%%%%%%%%%%%%%%%%%%%%%%%%%%%%%%%%%%%%%%%%%%%%%%%%%%%
%
\begin{abstract}
Using extensive numerical simulations, we investigate the flow
behaviour of a model glass-forming binary mixture whose constituent
particles have a large size ratio. The rheological response to
applied shear is studied in the regime where the larger species are
spatially predominant. We demonstrate that the macroscopic rigidity
that emerges with increasing density occurs in the regime where the
larger species undergo a glass transition while the smaller species
continue to be highly diffusive. We analyse the interplay between
the timescale imposed by the shear and the quiescent relaxation
dynamics of the two species to provide a microscopic insight into
the observed rheological response. Finally, by tuning the composition
of the mixture, we illustrate that the systematic insertion of the
smaller particles affects the rheology by lowering of viscosity of
the system.
\end{abstract}                   
%
%%%%%%%%%%%%%%%%%%%%%%%%%%%%%%%%%%%%%%%%%%%%%%%%%%%%%%%%%%%%%%%%%%%%%%%%%%%%
\maketitle

\section{I. Introduction}
The rheological properties of soft glassy materials (colloids, gels,
foams, emulsions, granular matter etc.) \cite{larson, yodhreview,
ludovicRMP2017, joshipetekidis, coussot2014yield, nicolas2018deformation,
rodney2011modeling} are very important for a wide range of
applications in our daily lives as well as in industries, and also
lead to various natural phenomena. A characteristic feature of soft
glassy materials is the existence of a yield stress, i.e.~a threshold
stress value which needs to be overcome to make the material flow.
Once flowing, soft glasses typically behave as shear-thinning
materials. Understanding the processes that lead to yielding and
flow from a microscopic perspective is fundamental to our physical
knowledge of the properties of soft glasses and could be utilized
for the development of soft glasses that are functionalized towards
specific applications.

Many soft glass systems consist of particles having a wide range
of sizes. This is an important issue with respect to their mechanical
properties. However, in experiments \cite{pusey2002, besseling2007,
schall2007, chikkadi12} and simulations \cite{yamamoto98, ludovic2002,
varnik2004, gauravPRE15, gauravJRheo15, anshul17, falkLanger98,
sollich2012, sollich2013} on the rheology of soft glasses, the focus
has been on model systems with a moderate size dispersity. Here,
the size dispersity has essentially been introduced to prevent the
system from crystallization. However, recent studies \cite{poon2015,
laurati17} have revealed that the degree of polydispersity has a
strong impact on the glassy dynamics. In systems with a large
polydispersity, there is a time-scale separation between the
structural relaxation of large and small particles that leads to
strong dynamical heterogeneities. How such a dispersity in particle
size affects the rheological response is not well understood and
needs to be further explored.

To address this issue, it is useful to consider disparate-sized
binary mixtures. In the quiescent state, such systems are known to
exhibit a complex phase behaviour \cite{dhont95} and a wealth of
fascinating phenomena with respect to their glassy dynamics
\cite{colmenero06, colmenero06pre, juergen2009, laurati15, laurati19}.
In the regime where the large particles predominantly occupy volume,
there is a distinct difference in the dynamics of large and small
particles as the density or packing fraction is increased. The
larger species are observed to undergo a glass transition first,
while the smaller species remain highly diffusive. Subsequently,
at much higher density, the smaller species reach dynamical arrest,
with their motion reminiscent to the localization dynamics of tracer
particles in an amorphous matrix of frozen-in soft spheres
\cite{skinner2013, schnyder2017, schnyder2018}.

Only recently, there has been systematic experimental investigations
of the transient and steady-state rheological response of binary
colloidal mixtures having large size disparity between the constituent
particles \cite{sentjabrskaja13,egelhaaf13, sentjabrskaja14,
sentjabrskaja18, sentjabrskaja19}.  These investigations, typically
done at a fixed volume fraction and using different shear protocols,
reveal that the shear response changes varying the concentration
of the smaller species.  As long as the larger particles occupy
more volume one observes a softening of the system, while it hardens
when the volume occupied by the smaller species becomes predominant.
This changing rheological response has been attributed to the change
in local packing structures as the composition of the mixture is
varied.

In this work, we report an extensive numerical study of the rheological
behaviour of a model glass-forming binary mixture having large
bidispersity, considering the regime where the larger particles
predominantly occupy volume. We consider a model system where the
two constituent species undergo separate dynamical arrests at vastly
different densities \cite{juergen2009}. We probe how the timescale
introduced by the external shear interacts with the relaxation
dynamics of the constituent species, which is then manifested in
the flow behaviour.  The main finding is that the measured rheological
curves reveal the onset of macroscopic rigidity in the density
regime where the larger particles undergo a mode coupling transition
in the quiescent state.  Importantly, when this rigidity is observed
at macro-scale, the smaller species continue to express a fluid-like
dynamics that is not affected by the external shear. Thus, there
is a macroscopic solid-like response even though internally there
still exists a fast dynamic species. Only at much larger densities,
the external shear is seen to affect the dynamics of the smaller
particles. However, these fast particles do influence the rheology
near the regime where rigidity sets in; we observe that the systematic
insertion of the smaller particles into the mixture leads to a
lowering of the viscosity, i.e.~to the softening of the material,
consistent with earlier experimental observations.

The paper is organised as follows. After the introductory discussion
in Section I, the model system and the numerical methods are discussed
in Section II. In Section III, we discuss our findings, with
subsections on (i) recapitulation of equilibrium results that have
been reported previously, (ii) elaboration of macroscopic and
microscopic analysis of the response to applied shear, and (iii)
illustration of how the smaller species influence the rheological
response. Finally, in Section IV, we have a concluding discussion.

\section{II. Model and computer simulation}
The model glass-former that we consider for our study is a $50:50$
binary mixture of repulsive particles whose quiescent dynamics has
been well studied \cite{juergen2009}. The bigger particles (labelled
A) have a slight polydispersity in size to avoid crystallization;
their diameters $d_{\rm A}$ are sampled from a uniform distribution,
i.e., $d_{\rm A} \in [0.85,1.15]$. The diameter of the smaller
particles (labelled B) are fixed to $d_{\rm B} = 0.35$. The average
size ratio of A and B particles is $\langle d_{\rm A} \rangle/d_{\rm
B} \approx 2.85$ where $\langle d_{\rm A} \rangle/ \approx 1$.
Details about the model interactions between these constituent
particles, as well as the units for length,
energy, and time  can be found in Refs.~\cite{juergen2009, vinay}. 

%Length,
%energy, and time are measured in units of $\langle d_{\rm A} \rangle$,
%$\epsilon_{\rm AA}$ and $[m \langle d_{\rm A}\rangle^2/\epsilon]^{1/2}$
%respectively.

We study the rheological behaviour of this model system using
molecular dynamics (MD) simulation via LAMMPS \cite{lammps}. We
consider a system consisting of $N = 2000$ particles in a
three-dimensional box with periodic boundary conditions at constant
temperature $T = 2/3$, which is maintained via a dissipative particle
dynamics (DPD) thermostat \cite{dpd}. The time step used for the
simulations is $dt = 0.00075$. Initial states at different densities
are sampled from our study on the quiescent behaviour of this
system; see Ref.~\cite{vinay}. Apart from the 50:50 composition,
we also study a couple of other compositions, viz.~100:0 and 75:25,
in order to understand how the inclusion of B species changes the
system's response to an external shear.

\begin{figure}[t]
\includegraphics[width=8cm]{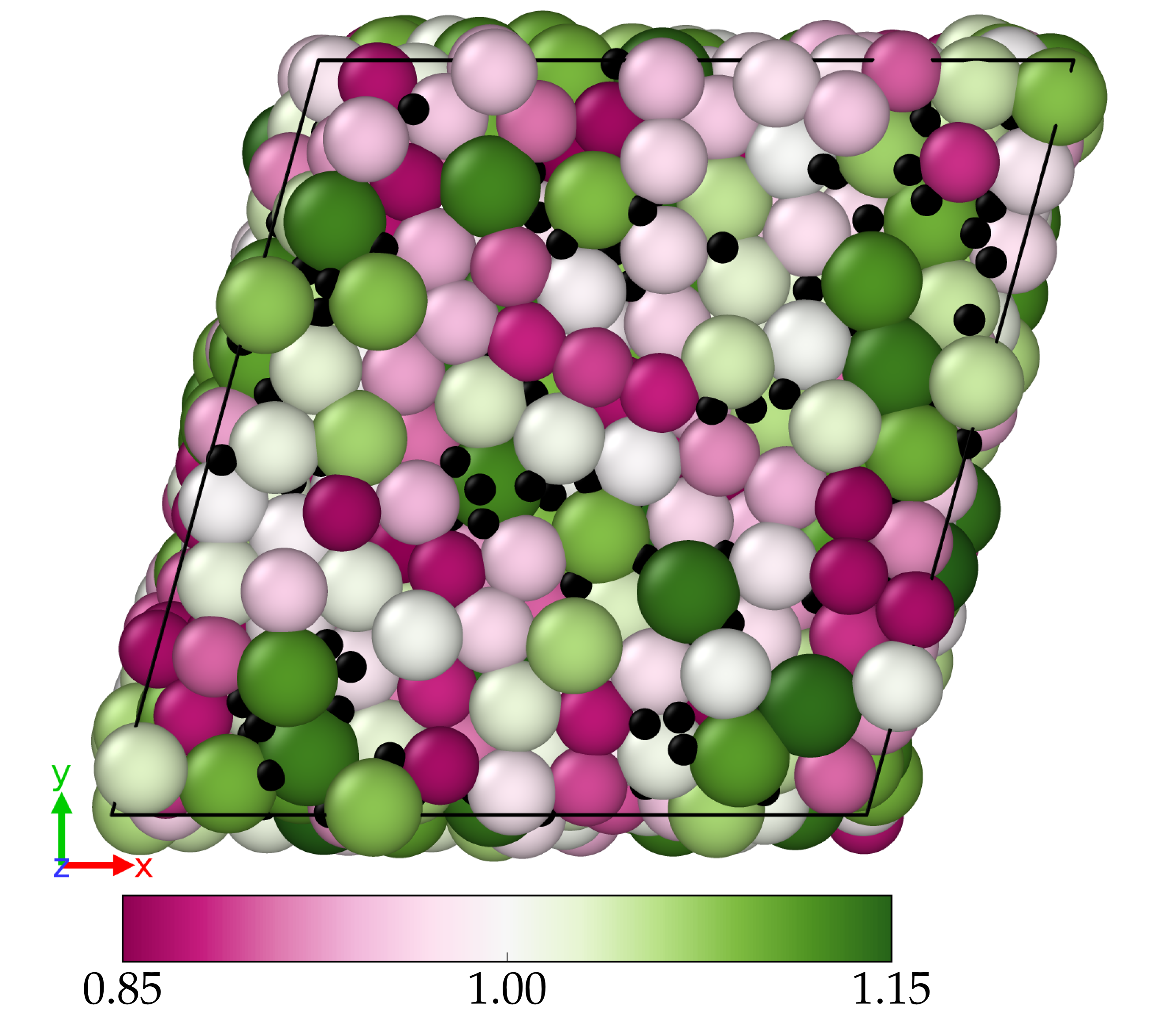}
\caption{{\em Snapshot of the sheared binary mixture.} The color
panel represents the diameter of the polydisperse larger species
(A).  Smaller particles (B) are shown in black. The shear is applied
in the $x$-direction along the $xy$ plane. \label{fig2}}
\end{figure}

To investigate the rheological response of the system, we impose
shear along the $xy$ plane in the $x$-direction (see Fig.~\ref{fig2})
using different shear rates ($\dot{\gamma}$) ranging between
$1.5{\times}10^{-6}$ and $10^{-3}$. Lees-Edwards boundary
conditions are utilised during the shearing. The rheological behaviour
is studied for a range of densities varying from very small, viz.
$\rho_{\rm A}=1.05$ to very large $\rho_{\rm A}=1.75$, where $\rho_{\rm A}$
is the partial density of A species.  Note that all reported densities will be in terms of $\rho_A$, since
we will be eventually probing how changing $\rho_{\rm B}$ influences the
rheology at fixed $\rho_{\rm A}$, as discussed above. A snapshot of the
sheared system is shown in Fig.~\ref{fig2}. As is evident, the larger
species indeed populate most of the volume and the smaller species
dot the intervening space. In this study, we inquire how the
microscopic dynamics of these two species are manifested in the
macro-rheology during the shear process.

\section{III. Results}
\subsection{Quiescent dynamics}
\begin{figure}[]
\includegraphics[width=8cm]{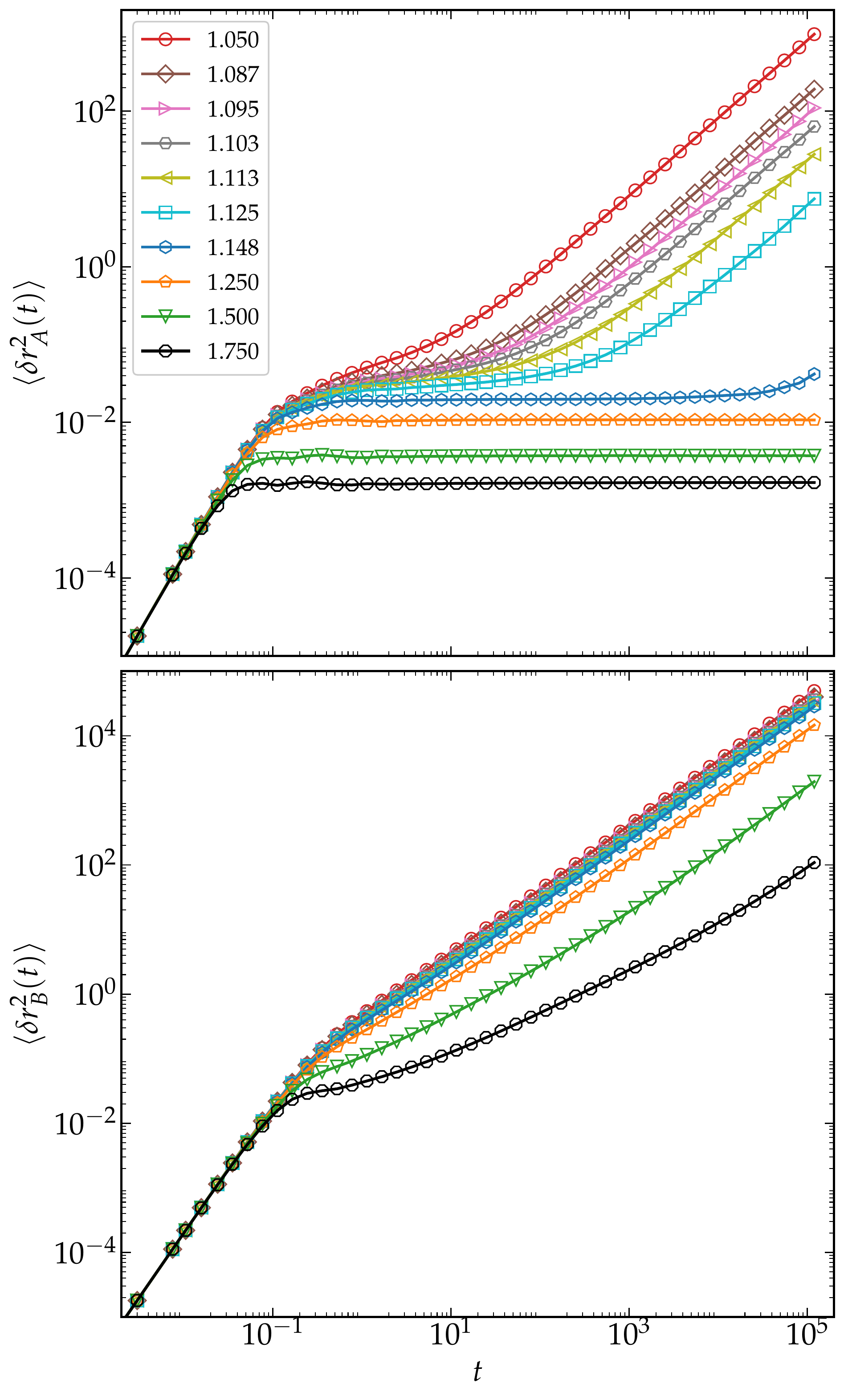}
\caption{{\em Quiescent dynamics.}  Mean-squared displacement of
bigger (upper panel) and smaller (lower panel) species of the
unperturbed system i.e., in the absence of shear for different
partial densities of bigger species. Densities are labelled in terms
of partial density of A species $\rho_{\rm A}$. \label{fig1}}
\end{figure}
As an introductory step, we summarize the dynamical behaviour
observed by scanning across densities and monitoring the time
evolution of the mean squared displacement (MSD) of the two species;
see Fig.~\ref{fig1}. As reported earlier in Ref.~\cite{juergen2009},
the dynamics of the A species falls out of equilibrium around the
critical density of mode coupling theory, $\rho_{\rm A}^{\rm
MCT}=1.115$).  In the same density regime, the B species remain
diffusive and undergo a dynamical arrest at a much higher density
\cite{juergen2009}. Or in other words, we observe two separate
onsets of glassiness in the sub-populations, due to the large size
ratio of the two species. Thus, at any density, the dynamics of the
B species are faster than that of the A species, i.e., the relaxation
timescales are relatively much slower. This large separation of
timescales plays an important role in the rheological response,
when a shear is introduced by driving the system at different
shear-rates. The shear imposes a new timescale into the system and
its interplay with the intrinsic timescale leads to the diversity
in rheological response.

We also note here that for the A particles the MSD is measured in
their center-of-mass reference frame. With this, we avoid finite-size
effects in the high density regime, i.e.~for $\rho > \rho_{\rm
A}^{\rm MCT}$, occurring due to a relatively large center-of-mass
diffusion coefficient of the A species for the considered relatively
small system sizes.  As a consequence, the A particles move
collectively with their center of mass, despite their dynamical
arrest with respect to particle rearrangements.  For further
discussion on this, see Ref.~\cite{vinay}.

\begin{figure}[]
\includegraphics[width=8cm]{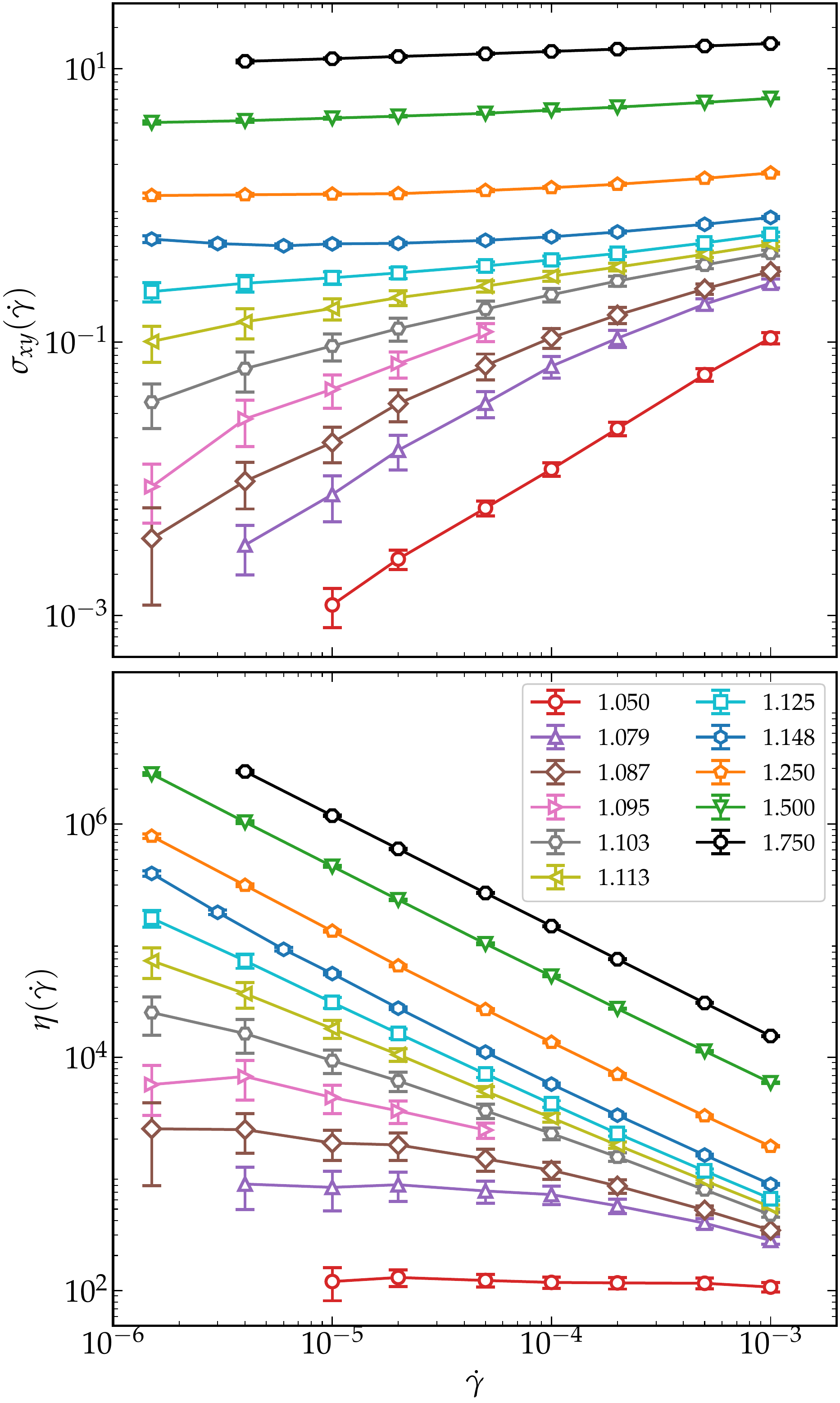}
\caption{{\em Rheological response.} Variation of steady-state shear
stress $\sigma_{xy}$ (top) and corresponding viscosity $\eta$
(bottom) with applied shear-rate $\dot{\gamma}$, at different partial
densities ($\rho_{\rm A}$) of bigger species as marked. \label{fig4}}
\end{figure}
\begin{figure}[]
\includegraphics[width=8cm]{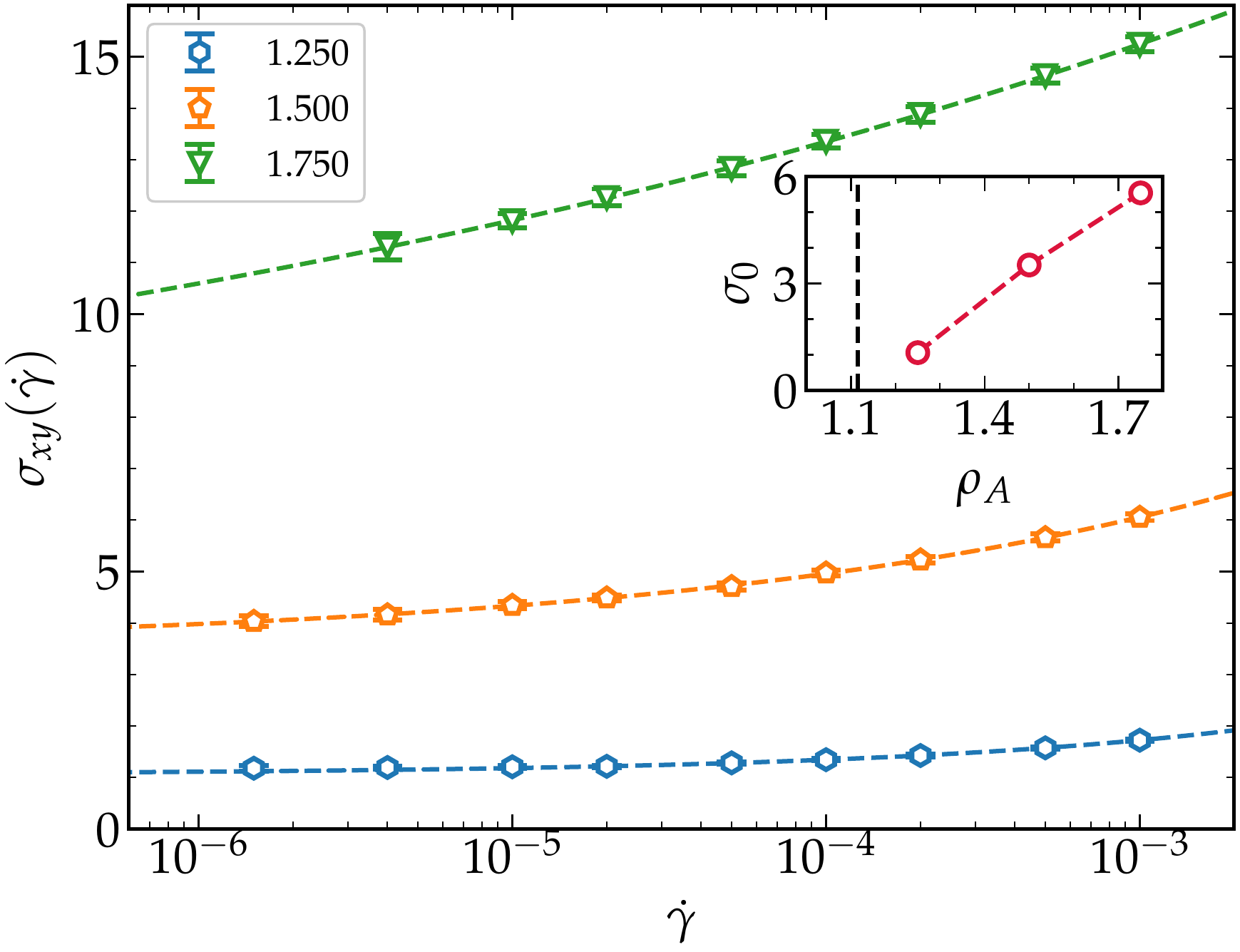}
\caption{{\em Herschel-Bulkley fits.} Flow curves, i.e., shear
stress vs.~strain-rate at different partial densities ($\rho_{\rm A}$) of
bigger species as marked, shown with corresponding Herschel-Bulkley
fit ($\sigma = \sigma_0+K\dot{\gamma}^n$) using dotted line. Fit
estimates are $\sigma_0 = 1.06, 3.52, 5.54$, $K = 8.33, 13.92,
18.63$ and $n = 0.36, 0.25, 0.09$ for $\rho_{\rm A} = 1.25, 1.50, 1.75$.
(Inset) Variation of estimated value of yield stress $\sigma_0$
with $\rho_{\rm A}$. Dotted vertical line marks $\rho_{\rm A}^{\rm
MCT}$. \label{fig44}}
\end{figure}
\subsection{Steady-state macroscopic shear response}
Next, we discuss the rheological response of the system over a range
of densities and shear rates. A snapshot of the system during shear
is shown in Fig.~\ref{fig2}, with the large size disparity of the
two species distinctly visible.

Since the shear is imposed along the $xy$ plane, we measure the
corresponding shear stress $\sigma_{xy}$ using the following
Irving-Kirkwood expression: $\sigma_{xy} = \langle \frac{1}{V}
\sum_{\alpha\beta} f^{x}_{\alpha\beta} r^{y} \rangle$, where
$f^{x}_{\alpha\beta}$ is the $x$-component of the force with respect
to a pair of particles and $r^y$ is the $y$-component of the distance
vector between two particles labelled $\alpha$ and $\beta$ which
could belong to either of the species A and B. $V$ is the total
volume of the simulated system.  $\langle\cdot\rangle$ corresponds
to averaging over states sampled in steady state.

At each density and imposed shear-rate, we obtain the steady state by
shearing the system to large strains and then ensuring that the
measured shear stress $\sigma_{xy}$ fluctuates around a steady mean
value. The data gathered for the variation of shear-stress with
imposed shear rate, i.e., the flow curve, for the wide range of
densities is shown in the top panel of Fig.~\ref{fig4}. And, the
corresponding viscosity defined by $\eta (\dot{\gamma}) =
\sigma_{xy}(\dot{\gamma})/\dot{\gamma}$ is shown in the bottom panel
of Fig.~\ref{fig4}.

The main findings from the rheological data are the following. At
small densities ($\rho_{\rm A} < 1.09$), a Newtonian regime, i.e., $\eta
(\dot{\gamma})$ is constant, is observed at small shear-rates. At
larger shear rates, shear thinning is observed, i.e., the viscosity
decreases with increasing $\dot{\gamma}$. The Newtonian regime
shrinks with increasing density, i.e., pushed to smaller and smaller
shear rates. These are distinct characteristics of complex liquids.
As the density is increased, an apparent yield stress becomes visible
at $\rho_{\rm A} \approx 1.125$, with $\sigma_{xy}$ becoming flat with
decreasing $\dot{\gamma}$. This implies that there is an onset of
macroscopic rigidity for the binary mixture. Around this density
regime, there is also a change in behaviour of $\eta (\dot{\gamma})$
-- within the window of small shear rates, the viscosity continues
to be an increasing function, i.e., there is no tendency to change
curvature towards plateauing out. With further increase of density,
the apparent yield stress of the system, viz.~$\sigma_{xy}(\dot{\gamma}
\rightarrow 0)$, increases steadily, which is characteristic to
amorphous solids. One can fit the variation of stress with shear rate
in this large density regime using the Herschel-Bulkley (HB) function,
$\sigma = \sigma_0+K\dot{\gamma}^n$, where $\sigma_0$ is the estimated
yield stress, $K$ is a constant and $n$ is the HB exponent. The
fits to the measured data for different $\rho_{\rm A}$ are shown in
Fig.~\ref{fig44} and the values of the fit parameters are listed in
the caption. In the inset of Fig.~\ref{fig44}, we display the
emergence and increase of the estimated yield stress $\sigma_0$
with density, beyond $\rho_{\rm A}^{\rm MCT}$ .

Overall, such variation of rheological flow curves, viz.~the emergence
of an yield stress with the variation in a control parameter 
(e.g.~temperature or packing fraction or density) are very similar to the
ones obtained in other soft-sphere glassy systems \cite{ludovic2002,
sollich2012, sollich2013, golkia2020}.

\begin{figure*}[]
\includegraphics[width=16cm]{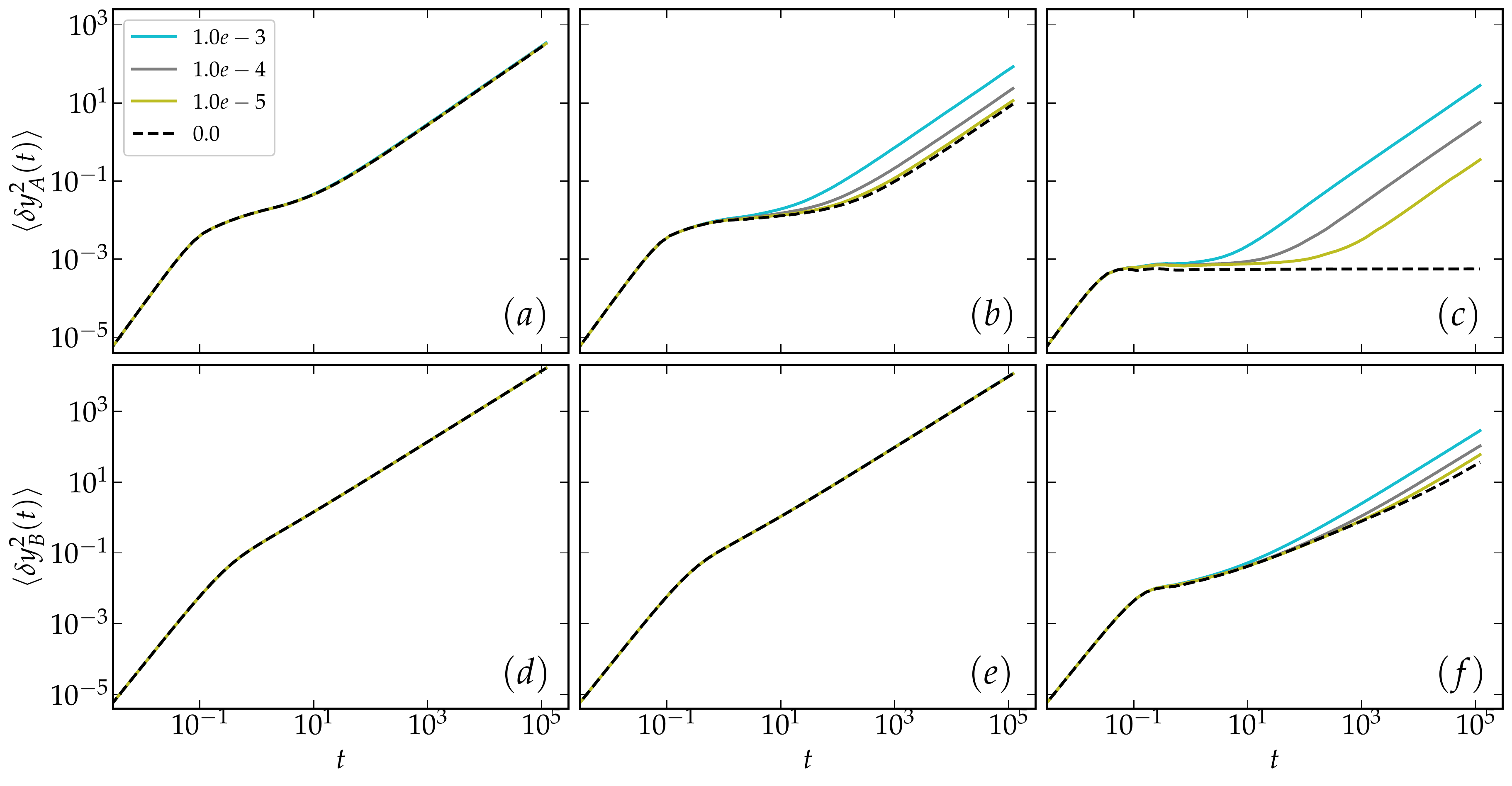}
\includegraphics[width=16cm]{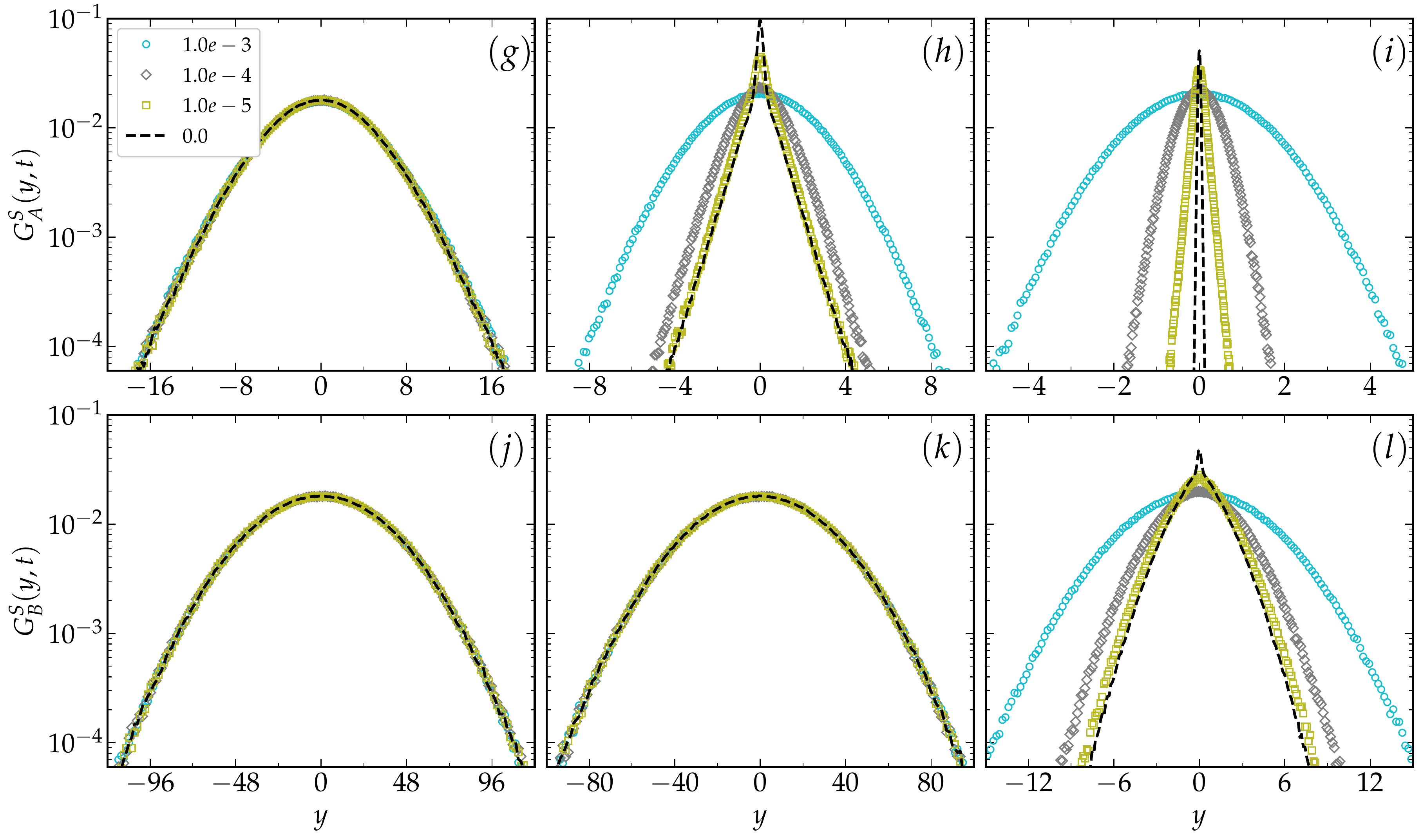}
\caption{{\em Microscopic dynamics under applied shear}: Mean squared
displacement of bigger (a-c) and smaller (d-f) species in the
presence (with solid lines) and in the absence (with dotted lines)
of shear, measured in the vorticity direction. Corresponding self
part of van Hove function, $G_s(y,t)$ of bigger species (g-i) and
smaller (j-l) species in the presence (with symbols) and absence
(dotted lines) of shear, at $t=8075.4$, also measured in the
vorticity direction. In both cases, first column represents system
at partial density $\rho_{\rm A} = 1.0500$, middle at $\rho_{\rm A} = 1.1125$ and
last at $\rho_{\rm A} = 1.7500$. \label{fig5}}
\end{figure*}
\subsection{Steady-state microscopic dynamics}
In order to further analyse the rheological response, we study the
microscopic dynamics of the system in steady state. In the vorticity
direction, i.e., normal to the shear plane, we compute the MSD
for the A and B species, separately, as well
as the self part of the van Hove function of which MSD is the second
moment.  The van Hove Function, $G_s(r,t)$ is measured at a late
time ($t=8075.4$) in each case. The corresponding aggregated plots
showing the variation of the dynamics with changing density are
shown in Fig.~\ref{fig5}, for MSD and $G_s(r,t)$, for a range of
imposed shear rates. All measurements are done in the center-of-mass
reference of respective species, to remove the effects resulting from finite
center-of-mass diffusion in the glassy regime of A species.

At the smaller density ($\rho_{\rm A} = 1.0500$), we do not observe any
variation in the dynamics with imposed shear rate, for both A and
B species, and the MSD curves as well as the $G_s(r,t)$ data are
indistinguishable between the equilibrium and non-equilibrium
regimes. This corresponds to the linear response regime, which is
characterised by the Newtonian viscosity, see Fig.~\ref{fig5} (a),
(d) for the MSD data and the corresponding data for $G_s(r,t)$ in
Fig.~\ref{fig5} (g), (j).

If we now consider the higher density, $\rho_{\rm A} = 1.1125$,  which
is in the vicinity of $\rho_{\rm A}^{\rm MCT}$, we observe that for the B
species, the indistinguishability between the equilibrium and
non-equilibrium data for MSD and $G_s(r,t)$ continue, i.e., the
external shear is having no effect; see Fig.~\ref{fig5} (e), (k).
However, for the A species, the shear introduces deviation from
equilibrium behaviour, with an enhancement in long-time diffusive
dynamics with increasing shear rate, as is visible in the MSD data
(Fig.~\ref{fig5} (b)) which is also evident in the distinct change
in the shape of the corresponding $G_s(r,t)$ (Fig.~\ref{fig5}(h)).
Thus, at this density, the imposed shear rate corresponds to a
timescale which is faster than the equilibrium relaxation timescale
of the A species, which is not the case for B species. Note that,
for this density, an apparent macroscopic yield stress is visible.
Thus, we can conclude that the apparent emergence of macroscopic
rigidity and subsequent yielding under shear is a consequence of
the interplay between the external drive and the structure formed
by the A species, while the B species remain oblivious to this
process and thus seems inconsequential to the overall rheological
response, i.e., variation of viscosity with shear rate in this
density regime.

However, if we go to even larger densities, $\rho_{\rm A}=1.7500$, the
imposed shear affects the dynamics of both the A and B species; see
Fig.~\ref{fig5} (c), (f), (i), (l); for both species, enhanced
diffusion is observed with increasing shear rate. Thus, the imposed
timescales via the shear are faster compared to the equilibrium
relaxation timescale for both species, even though there is still
a large separation in dynamical timescales in the quiescent state.

\begin{figure}[t]
\includegraphics[width=8cm]{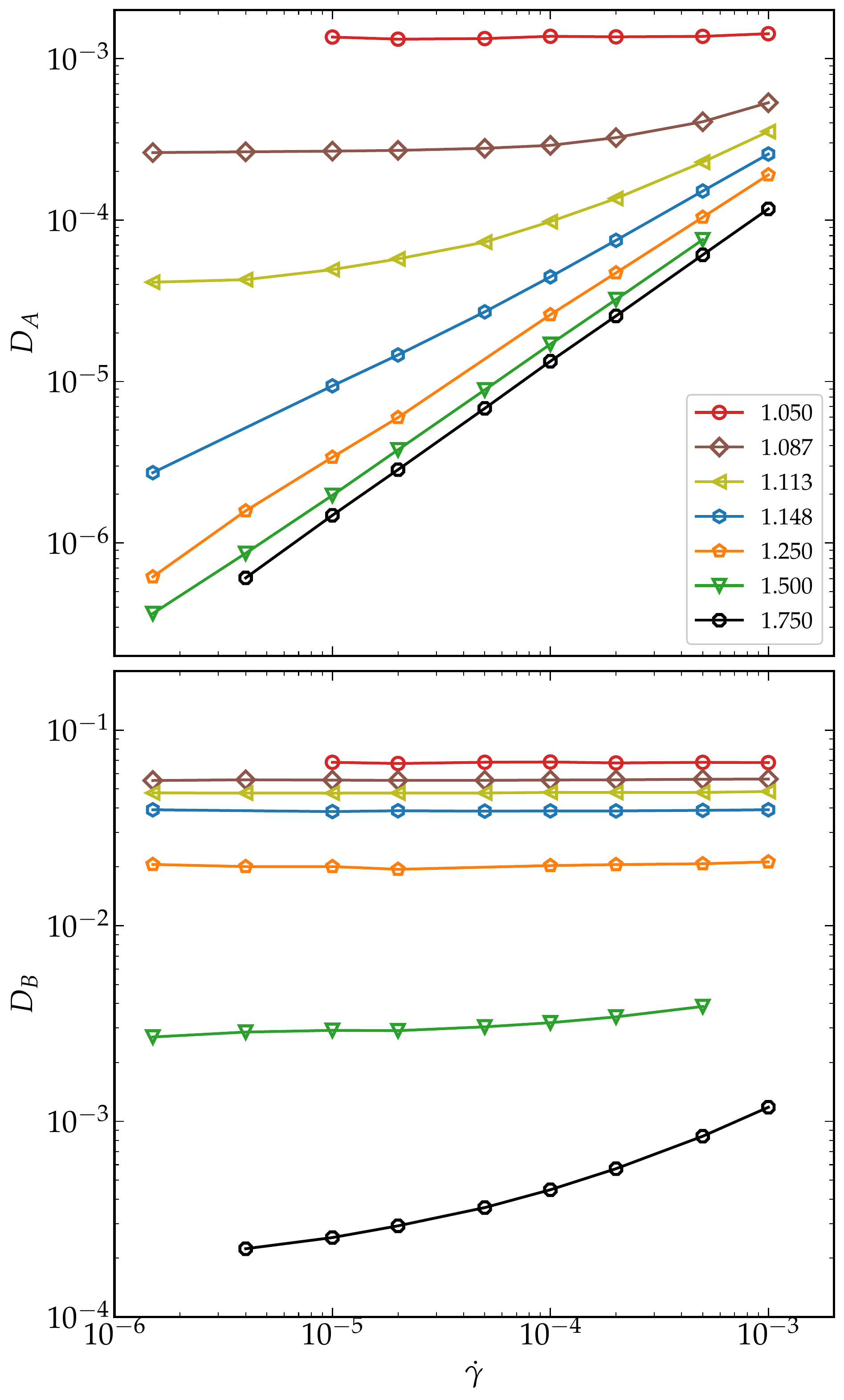}
\caption{{\em Diffusion coefficient in the presence of shear.}
Variation of diffusion coefficient of bigger (top) and smaller
(bottom) species with shear rate, measured in the vorticity direction,
at different partial densities of bigger species as marked. \label{fig7}}
\end{figure}

We can extract a diffusion coefficient from the long-time diffusive
dynamics observed in the MSD data, for the various imposed shear rates
at different densities. The shear-rate dependence of the diffusion
constant, for the range of densities explored, is shown in
Fig.~\ref{fig7}, for both A and B species.

At small densities, the diffusion coefficient for both A and B
species has no shear-rate dependence, which is just a reflection
of the fact that the dynamics remains unaffected by the shear, as
discussed above. This behaviour continues for B species up to very
large densities where the shear starts to have an effect. Also, in
this density regime, the diffusion coefficient for A species has a
power-law dependence on shear rate and appears to vanish at
$\dot{\gamma} \rightarrow 0$, which is typical to the shear response
of amorphous systems. In the intermediate density regime, the
diffusion coefficient also decreases with decreasing shear rate,
but deviates from a power-law behaviour and seems to reach a constant
at vanishing shear rates. This would imply that there is a diffusive
behaviour in the absence of shear at these densities. This behaviour
is observed around the density where the onset of glassiness occurs
for the A species, viz.~in the vicinity of $\rho_{\rm A}^{\rm MCT}$. We also
note that for B species a similar behaviour is observed at the largest
density that we have explored, implying that in the absence of
shear, the smaller particles would be undergoing equilibrium diffusive
dynamics in that density regime.

To summarise, from the rheological response of the binary mixture,
we infer that beyond $\rho_{\rm MCT}$, an apparent yield stress emerges
which is linked to the glassiness undergone by the A species. On
contrary, the B species continue to be very mobile inside the matrix
formed by the A species and their dynamics does not couple to the
shear, and thus their contribution to the system's shear response
is minimal. Only at much higher densities, the yielding process
involves the collective interplay of both A and B species.

\begin{figure}[htb]
\includegraphics[width=8cm]{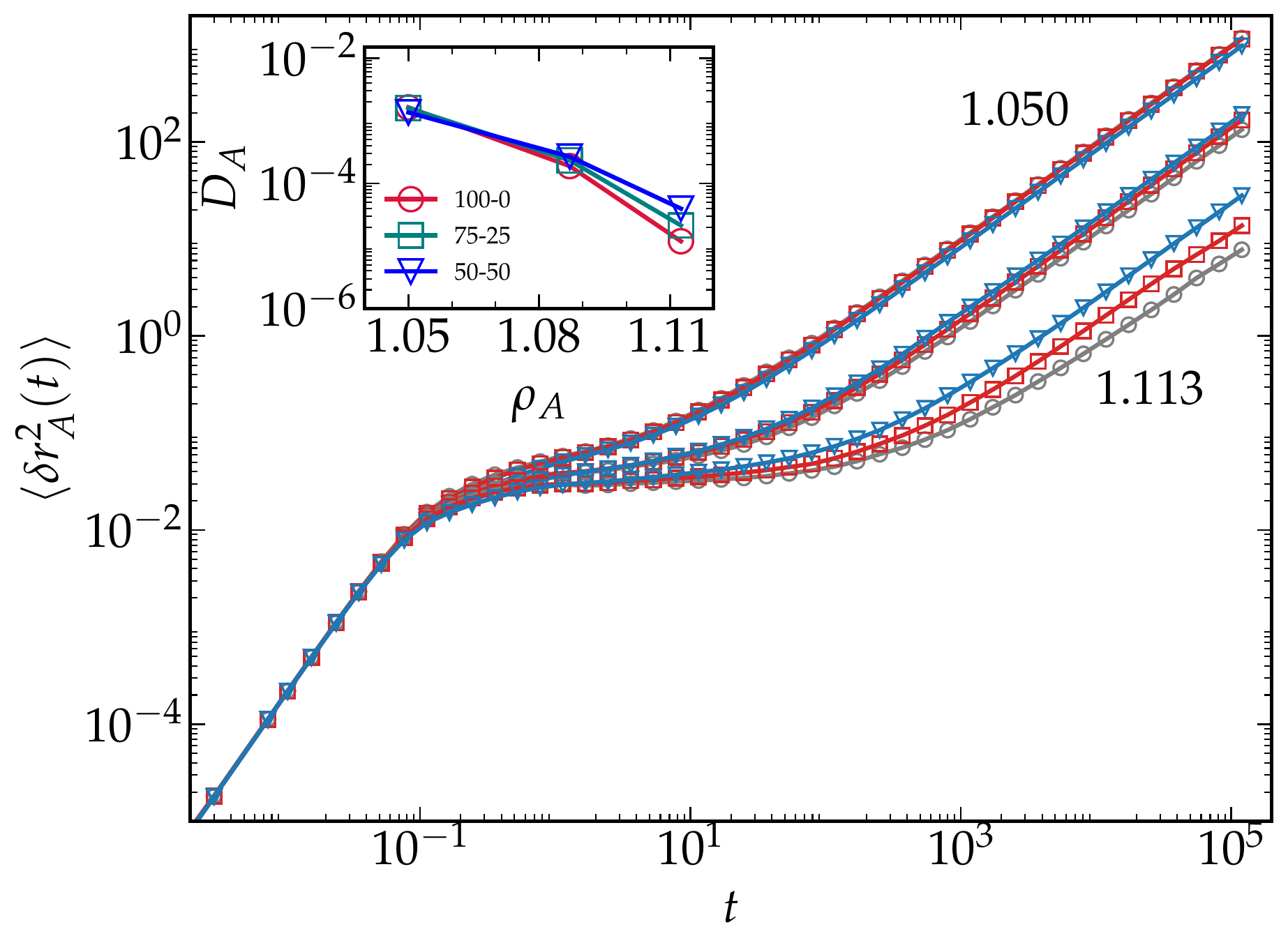}
\includegraphics[width=8cm]{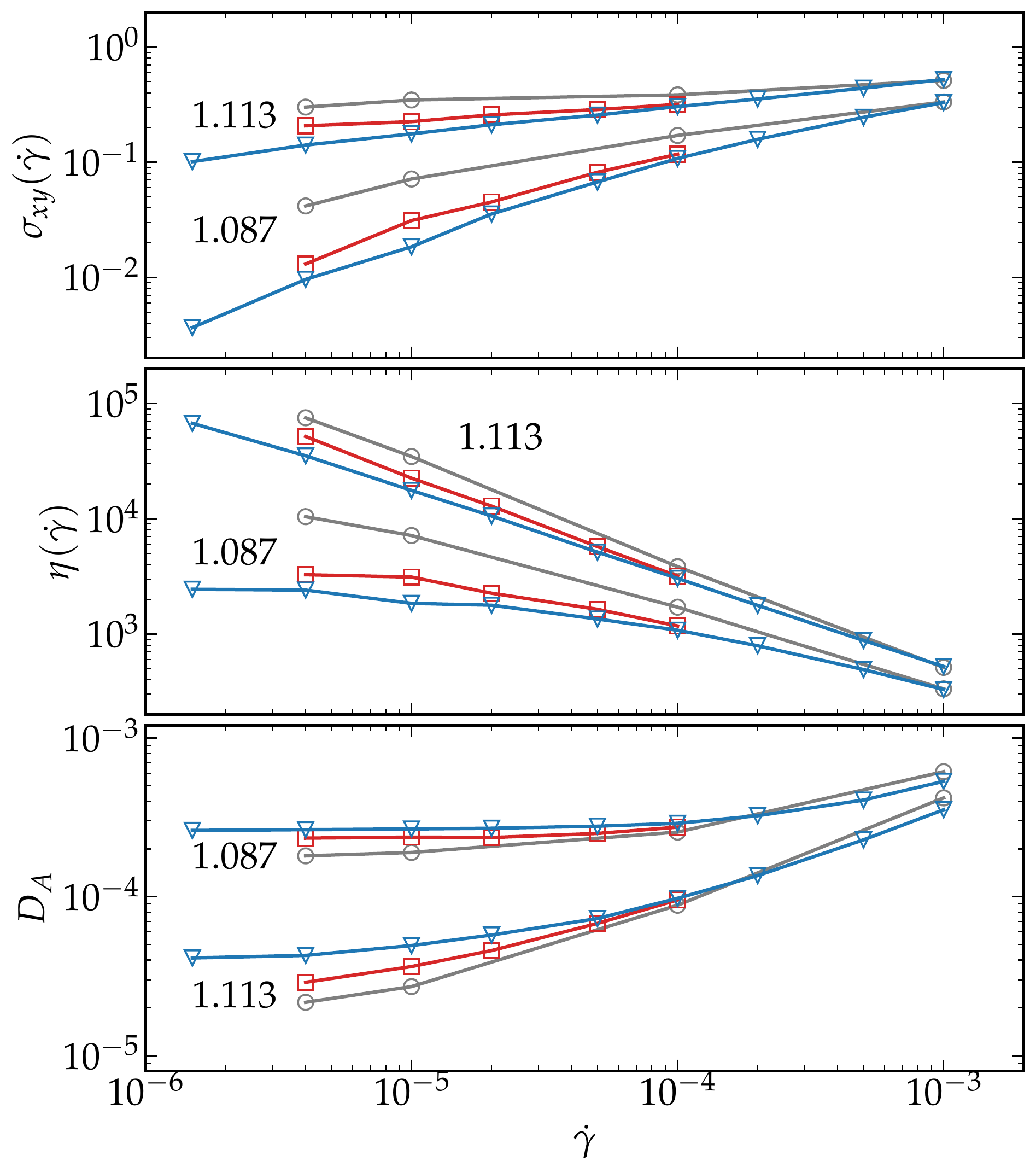}
\caption{{\em Effect of varying composition}. (Top) {\em  Quiescent
behavior.} MSD measured for the A species, for different composition
of the mixture (circle: 100-0, square: 75-25, triangle: 50-50).(Inset)
Variation of larger particle diffusion coefficient with partial
density $\rho_{\rm A}$ for different composition of the system. (Bottom)
{\em Rheological response.} Shear stress and viscosity as a function
of shear-rate for different composition of the mixture.  And,
diffusion coefficient versus shear rate for the same compositions.
\label{fig9}}
\end{figure}
\subsection{Role of smaller species: varying the composition}
We now address the question of what role do the B species play in
the rheological response.  We address this questions by varying the
composition of the mixture, and even considering the situation where
the B species are completely absent.

We begin by first examining how the equilibrium behaviour of the
system is influenced by the presence or absence of the B species.
In the top panel of Fig.~\ref{fig9}, we show the MSD data for the A species
for varying A:B compositions, viz.~100:0, 75:25 and 50:50 for three
different $\rho_{\rm A}$. We note that at small densities, there is no
variation in the dynamics, but at relatively larger densities
($\rho_{\rm A} = 1.113$) there is distinct variation with the long-time
dynamics becoming faster with increased presence of B species. This
is also reflected in the measured diffusion coefficients as shown
in the inset of the top panel of Fig.~\ref{fig9}. From this data, it
would imply that $\rho_{\rm A}^{\rm MCT}$ shifts to larger densities with the
increasing inclusion of B species. It is interesting to note that,
the insertion of the smaller B species has a very different effect
than insertion of bigger particles in a system of smaller particles.
In the latter case, the bigger particles act as pinning sites and
slow down the dynamics.

Next, we study how the rheological response changes, in the presence
or absence of the small particles. In the bottom panel of Fig.~\ref{fig9},
we do a comparison of the rheological curves for the different
compositions listed above. The effect of enhancement in equilibrium
dynamics due to the presence of B species is reflected in the
rheology data, especially in the intermediate density regime (as the
glassy regime of A species is approached), where the viscosity of
the system distinctly decreases at small shear rates with increasing
presence of B species. This effect is absent at small densities and
also at very large densities where the entire system becomes glassy.
Thus, in the intermediate density regime, the mobile B population
does have a softening effect on the overall shear response, an
effect which can be exploited in various applications.

Our findings are consistent with experiments involving colloidal
glasses with constituents having large size ratio, where a similar
softening of the material was observed with increasing insertion
of small particles till a 50:50 composition was reached
\cite{sentjabrskaja18, sentjabrskaja19}.

\section{IV. Conclusions and perspective}
In conclusion, we have studied the shear response of a glass forming
model binary mixture having a large size ratio between the constituent
species. The mixture having a composition of 50:50 corresponds to
the case where the larger species occupy most of the volume. For
such a system, it had earlier been demonstrated that as the overall
density is increased, the larger particles undergo dynamical arrest
first while the small particles continue to be diffusive within the
matrix formed by the large ones. In our study, we probe how the
external drive applied via a fixed shear rate, of varying magnitude,
influences this microscopic dynamics and how that manifests in the
observed steady-state rheological response.

The main observation comes via the measured shear stress, in steady
state, as a function of the imposed shear rate. At small densities,
the low shear-rate regime corresponds to that of a Newtonian fluid.
However, in the density regime where the larger species within the
mixture undergo a mode coupling dynamical transition in the quiescent
state, the shape of the flow curves change at vanishing shear rates
and an effective yield stress can clearly be estimated. Thus, there
is a macroscopic rigidity. If we now focus on the microscopic
dynamics, we observe that while the motion of the larger species
indeed get influenced by the external shear, the dynamics of the
smaller species remain completely unchanged vis-a-vis their motion
in the absence of shear, within the explored range of shear rates.
Thus, the macroscopic rigidity at vanishing shear rate is occurring
while there is a highly mobile population of the smaller species
within the system.

With increasing density, the yield stress increases as expected and
the diffusion coefficient of the larger species exhibit a distinct
power-law behaviour as a function of applied shear rate which is a
distinctive feature of glassy systems. Only at large densities, we
start to observe that the external shear rate is able to influence
the motion of the smaller species, i.e., the timescale imposed by
the applied shear rate is large enough to compete with the relaxation
timescale of the smaller species.

Finally, we probe the role of the smaller species in the rheological
response of the mixture, since the macroscopic rigidity emerges in
the system even though these particles remain highly mobile. We
observe that if we start with a system of just the large particles
and then start inserting the smaller species, i.e., systematically
change the mixture's composition, the rheological response changes
in the small shear-rate regime, viz.~the viscosity of the system
decreases. Thus, the presence of the smaller particles softens the
mixture's mechanical response, which is consistent with experimental
observation in similar colloidal mixtures. Note that similar
plasticizing effects have also been reported for polymeric systems
\cite{zaccarelli2005, kalathiprl2012} where the inclusion of very
small colloidal particles led to softening. Hence, this indicates
to a very generic mechanism at play in how such inclusions can
influence rheology.  We also note here that this behaviour is in
contrast to the situation where the inclusion of large particles
in a matrix of small particles can lead to hardening of the system,
i.e., the rheological behaviour of the mixture changes as the
composition is changed.

The next step should be to study the flow behaviour of glass forming
liquids where particles have large polydispersity in size
\cite{laurati17, ranjini2017}, to investigate how our findings on
the connection between microscopic dynamics and macroscopic mechanical
response get translated to more complex mixtures. Also, there is a
need to study how thermal fluctuations influence rheology in such
mixtures, by comparing the response in the athermal limit where
there have been some initial studies on probing the jamming behaviour,
the plasticity and also the shear response in asymmetric binary
mixtures \cite{ching10, atsushi21,luding20, pednekar2018, eric2015}.

\section{Acknowledgments}
We acknowledge the use of HPC facility at IMSc
Chennai for our computational work.
%

%

%%%%%%%%%%%%%%%%%%%%%%%%%%%%%%%%%%%%%%%%%%%%%%%%%%%%%%%%%%%%%%%%%%%%%%%%%%%%
\end{document}